\documentclass[aps,prx,twocolumn,floats,showpacs,superscriptaddress,nofootinbib]{revtex4-1}
\usepackage{graphicx,epsfig}
\usepackage{times,bbm}
\usepackage{graphics,dcolumn,bm,float}
\usepackage{amssymb,amsmath,rotate,color,amsfonts}
\usepackage[title,titletoc,toc]{appendix}
\usepackage{mathtools}
\usepackage{booktabs}
\usepackage{tcolorbox}
\usepackage[pagebackref=false,colorlinks,linkcolor=magenta,citecolor=blue,urlcolor=magenta]{hyperref}

\usepackage{wrapfig}
\usepackage{lipsum}
\usepackage{mwe}
\usepackage[mathlines]{lineno}
\usepackage{hyperref}
\usepackage{breqn}
\usepackage{bbold}
\usepackage{pgfplots}
\usepackage{float}
\usepackage{tkz-euclide}
\usepackage{braket}
\usepackage{physics}
\usepackage[caption=false]{subfig}
\usepackage[export]{adjustbox}
\usepackage{tikz}
\usetikzlibrary{through,calc}
\usetikzlibrary{positioning}

\newcommand{\be}{\begin{equation}}
\newcommand{\ee}{\end{equation}}

\newcommand{\bea}{\begin{eqnarray}}
\newcommand{\eea}{\end{eqnarray}}
\newcommand{\nn}{\nonumber}

\newcommand{\eps}{\varepsilon}

\newcommand{\pr}{\partial}

\newcommand{\bmt}{\left[\begin{matrix}}
\newcommand{\emt}{\end{matrix}\right]}

\begin{document}
\preprint{}
\title{Electrically charged Andreev modes in two-dimensional tilted Dirac cone systems}

\author{Z. Faraei}
\email{zahra.faraei@gmail.com}
\affiliation{Department of Physics$,$ Sharif University of  Technology$,$ Tehran 11155-9161$,$ Iran}
\affiliation{Abdus Salam ICTP$,$ Strada Costiera 11$,$ I-34151 Trieste$,$ Italy}

\author{S.A. Jafari}
\email{jafari@sharif.edu}
\affiliation{Department of Physics$,$ Sharif University of  Technology$,$ Tehran 11155-9161$,$ Iran}


\date{\today}

\begin{abstract}
In a graphene-based Josephson junction, the Andreev reflection can become specular which gives rise to propagating Andreev modes. 
These propagating Andreev modes are essentially charge neutral and therefore they transfer energy but not electric charge. 
One main result of this work is that when the Dirac theory of graphene is deformed into a tilted Dirac cone, the breaking of charge
conjugation symmetry of the Dirac equation renders the resulting Andreev modes electrically charged. We calculate an otherwise zero charge 
conductance arising solely from the tilt parameters $\vec\zeta=(\zeta_x,\zeta_y)$. The distinguishing feature of such a form of
charge transport from the charge transport caused by normal electrons is their dependence on the phase difference $\phi$ of the two
superconductors which can be experimentally extracted by employing a flux bias. Another result concerns the enhancement of Josephson current in a regime
where instead of propagating Andreev modes, localized Andreev levels are formed. In this regime, we find enhancement by orders of magnitude of the
Josephson current when the tilt parameter is brought closer and closer to $\zeta=1$ limit. We elucidate that, the enhancement is due to a 
combination of two effects: (i) enhancement of number of transmission channels by flattening of the band upon tilting to $\zeta\approx 1$, and
(ii) a non-trivial dependence on the angle $\theta$ of the the tilt vector $\vec\zeta$. 
\end{abstract}



\maketitle
\narrowtext

\section{Introduction}
\begin{figure}[t]
\centering
\includegraphics[width=9cm]{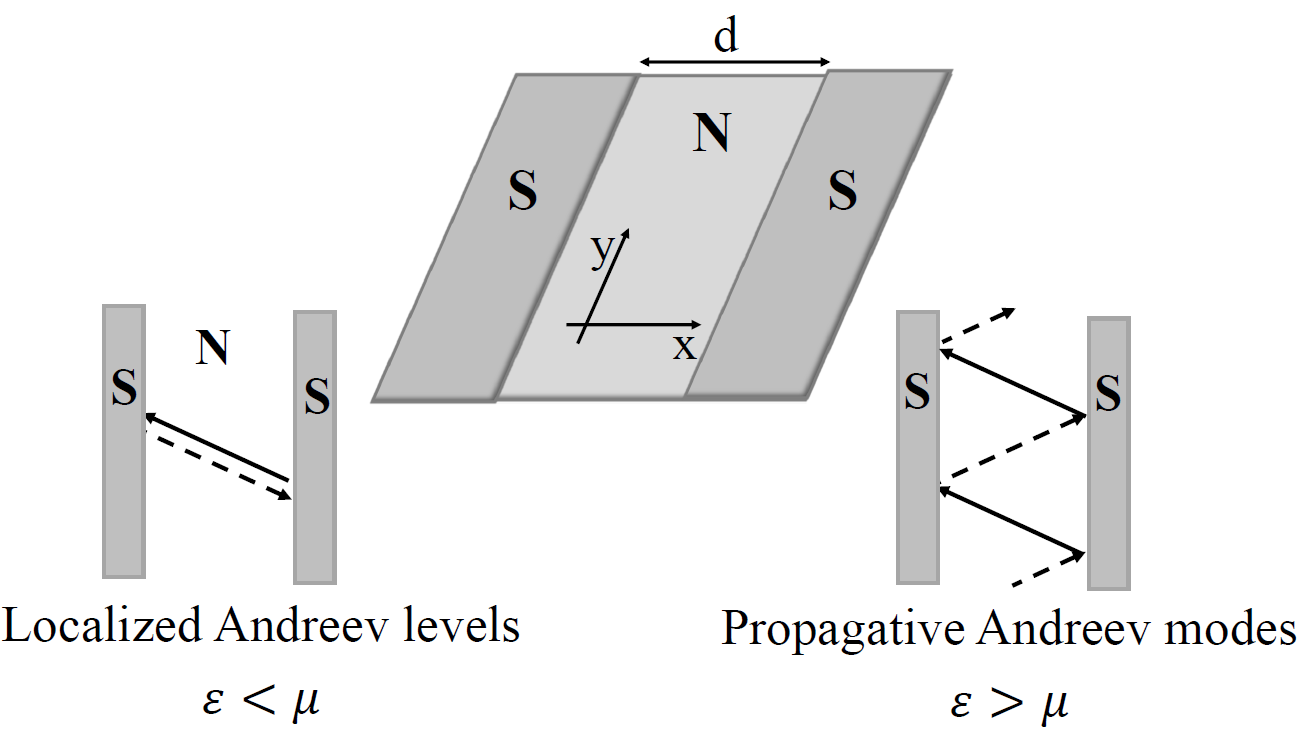}
\caption{Two types of Andreev reflection in Dirac material based SNS junctions~\cite{Beenakker_SAR}. The left panel shows the usual retro Andreev reflection creates the Andreev bound levels. The right panel illustrates the specular Andreev reflection which emerges from the linear  nature of the band structure of Dirac materials and is the reason for the formation of propagative Andreev modes. 
}
\label{AR.fig}
\end{figure}

Andreev introduced a particular type of reflections at the interface of  normal metal-superconductor (N-S) junctions which explains how the quasiparticles with energies below the superconducting gap can propagate into the superconducting region~\cite{Andreev} known today as Andreev reflection (AR). Electron-hole conversion implies the converting of the incident quasiparticle to its charge conjugated counterpart. In an ordinary NS junction~\cite{MET,BTK,Benistant,deGennes,ARsurvay,Kulik1969}, where the N region is a generic metal, the Andreev reflected particle is retro-reflected as illustrated in the left panel of Fig~\ref{AR.fig}. Each AR injects a Cooper pair into S region. Semiclassically, the quantization condition for Andreev bound state (ABS) corresponds to 
the situation where the total phase change of a quasiparticle in a closed path from one of the interfaces to the opposite one and then back to the first interface is $2\pi n$ 
with $n$ is an integer~\cite{nazarov2009}. These ABSs are responsible for the Josephson current which depends on the phase difference between the two superconductors~\cite{Josephson,Titov,Blackschaffer2008}.

Dirac materials are a class of materials where the two bands linearly touch at a Dirac node~\cite{Armitage2018,castro,geim}. 
Such touching points become the source of Berry curvature that not only heavily affects the 
semiclassical dynamics of electrons~\cite{NiuBerryPhase}, but also allows to encode the essential quantum anomaly
in odd space dimensions into a semiclassical description~\cite{StephanovChiralKinetic,SonYamomoto,SonSpivak2013}. 
The peculiar role of Dirac node is not limited to its the Berry curvature associated with it.
When it comes to the superconductors, this point will generate a new kind of AR. 
More than a decade ago, Beenakker realized that when the chemical potential $\mu$ of graphene is tuned near Dirac node, 
the specular AR (SAR) can also be possible~\cite{Beenakker_SAR}. As depicted in the right panel of Fig.~\ref{AR.fig},
in the SAR process, the reflection path resembles the reflection of a light ray from a mirror. This process has no
analog in the interface of non-Dirac conductors with superconductors. 
Indeed, the SAR is possible in Dirac materials in any space dimensions if the chemical potential is tuned close enough to charge
neutrality point. Because under such circumstances, the incident quasiparticles can choose their Andreev partner thorough an interband scattering which lead to 
specular Andreev reflection instead of retro Andreev reflection (RAR), the later being a consequence of intraband scattering processes. 
This is the direct consequence of the existence of a band touching touching point between the valence and conductance bands in 
band structure of Dirac materials~\cite{Shailos2007,Miao1530,Heersche,Yeyati2006}.   

The specular Andreev reflection creates a new kind of transport in SNS junctions as follows: In the right panel of Fig.~\ref{AR.fig},
sequences of repeated SARs have two effects: (i) Their first effect (similar to RAR processes) is to transfer a Cooper pair between the two S regions. 
This amounts to the ordinary Josephson current. (ii) They can additionally give rise to a propagating (dispersive) 
{\em Andreev mode} that will carry heat current along the N channel ($y$ axis in the middle panel of Fig.~\ref{AR.fig})~\cite{Andreevmodesgap}. 
Such a current is not a charge current due to the inherent neutral nature of Andreev modes, but it does transport energy. 
The unique feature of this form of AR-based transport is its dependence on phase difference of the superconducting leads. 
The dispersion of Andreev propagative modes in such a system has an excitation gap which is phase-dependent and this is the origin 
of phase-dependence in this type of energy current.

The root cause of the charge neutrality of the Andreev modes is the charge conjugation symmetry of the underlying Dirac equation (which is also
carried along all the way up to the in quantization condition of Andreev propagators in SNS setting). 
Breaking the charge conjugation symmetry is therefore expected to render the Andreev modes electrically charged. 
The natural way to achieve this is to tilt the Dirac 
cones~\cite{Beenakker2016KleinTunneling,Varykhalov2017,Lozovik,Soluyanov2015,Pyrialakos2017,Suzumura2006,Tajima2006,tohid,Cabra2013}~\footnote{
In a 2+1 dimensional Dirac theory relevant to present paper, the Pauli matrices $\sigma_x$ and $\sigma_y$ are used in the construction of the Hamiltonian $\vec\sigma .\vec p$. 
The only remaining matrices to deform this equation are $\sigma_z$ and the unit matrix $\sigma_0$. The matrix $\sigma_z$ will render the Dirac theory
massive which still has the charge conjugation symmetry. The only remaining option is to perturb it by matrix $\sigma_0$. At the linear oder, this perturbation
generates a tilt in the dispersion and breaks the charge conjugation symmetry of the original Dirac theory. 
}. 
In recent years, the remarkable effects of tilt in properties of two and three dimensional Dirac/Weyl materials have attracted much attention
~\cite{Bergholtz,Ogata2015,Aoki2011,Tohyama2009,Goerbig2014,Rostamzadeh}.
In addition to classical example of two-dimensional tilted Dirac cone in organic material~\cite{Katayama6,Suzumura2006,Suzumura2014}, 
and a certain structure of borophene called 8Pmmn borophene~\cite{Zhou2014,SaharTiltKink,Tarun2017},
it has been proposed that the partial hydrogenation of graphene can also give rise to anisotropic tilted Dirac cone~\cite{Ting2016}.
In fact tilting the Dirac/Weyl cones in any space dimensions, in addition to breaking the charge conjugation symmetry, will also mix energy and momentum space. 
Corresponding to this, the real space and time will be mixed. This point of view allows a mathematically neat and covariant formulation of the tilt in terms of a 
metric tensor~\cite{Saharmetric,JafariLorentz,tohid,Volovikblackhole} with interesting consequences~\cite{SaharPolariton}.

In this paper, we are interested in exploring the effect of the tilt deformation of the two-dimensional Dirac equation on the resulting Josephson current,
and the Andreev modes arising from SAR processes in Dirac material based SNS junction (see Fig.~\ref{AR.fig}). 
The major result of this paper is that the {\em neutral} Andreev mode of upright Dirac theory, acquires electric charge upon tilting and therefore a phase dependent
transport of charge along the N channel will be possible. This has no counterpart in non-tilted Dirac materials as it relies
on breaking of the charge conjugation symmetry of the Dirac equation by tilt. 
The paper is organized as follow: In section~\ref{formulation.sec} we introduce a minimal model to tilt the Dirac cone in two space dimensions
and obtain the quantization condition for Andreev reflected paths. In section~\ref{ChargedAM.sec} we study in detail the long junction limit of 
the quantization condition obtained in section~\ref{formulation.sec} and calculate the resulting thermoelectric transport coefficients
and their dependence on the tilt parameter $\vec\zeta$. In section~\ref{InfDOS.sec} we study the short junction limit that admits ABSs.
This allows us to obtain a detailed dependence of the Josephson current on the tilt vector $\vec\zeta$ where we find enhancement by orders
of magnitudes of the critical Josephson current by approaching the $\zeta=1$ limit. 

\section{Andreev reflection in tilted Dirac fermion systems}
\label{formulation.sec}
\begin{figure}[t]
\centering
\includegraphics[width=9cm]{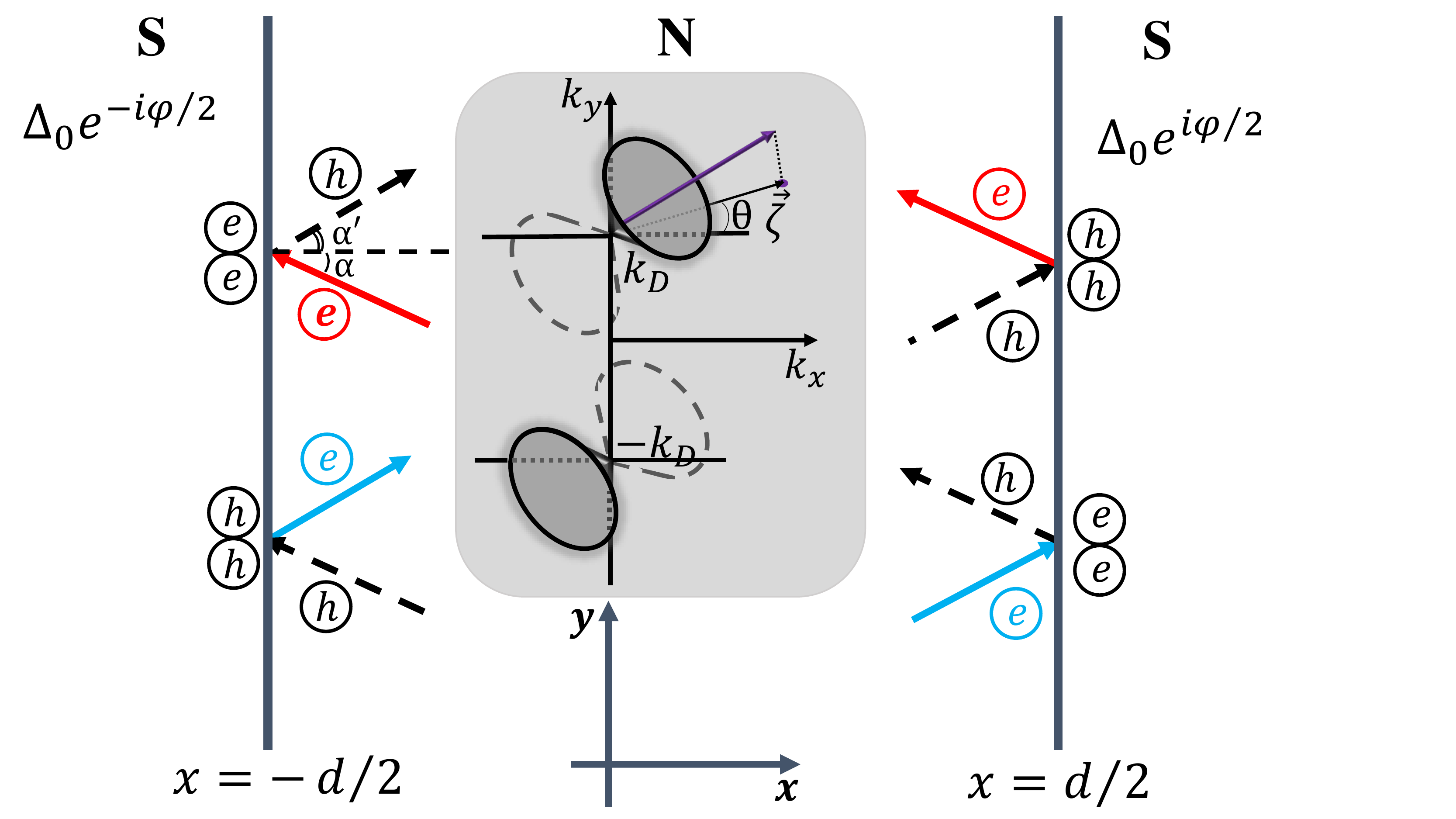}
\caption{Schematic representation of the specular Andreev reflections at each SN boundary. Right (left) moving
electrons are denoted by blue (red). The color code in this figure is the same as subsequent figures. 
The inset illustrates the tilting of the two Dirac cones in opposite directions along the $k_y$ axis. }
\label{model.fig}
\end{figure}

Consider a sheet of graphene in $xy$ plane as in Fig~\ref{model.fig}. The regions defined by $|x|>d/2$ are superconducting 
while the middle region, with $|x|<d/2$ is in the normal state of the 2D Dirac material. 
Let us deform the Dirac theory of graphene by a tilt term. As shown in the inset of Fig~\ref{model.fig}, the tilt is parameterized by a vector 
$\vec \zeta=(\zeta_x, \zeta_y)$ that corresponds to a tilt magnitude $\zeta=\sqrt{\zeta_x^2 + \zeta_y^2}$ along the angle $\theta=\tan^{-1}( \zeta_y/ \zeta_x)$ with respect to $k_x$ axis in momentum space. The value $\zeta=0$ corresponds to the upright Dirac cone, while the limit $\zeta=1$ shows a situation that the tilted Dirac cone tangentially touches 
the $k_xk_y$ plane along the direction specified by angle $\theta$ of the tilt vector. Other magnitudes of $\zeta$ correspond to the situations between these two limits. 
In order to satisfy the time reversal invariance of the entire system, the other Dirac cone on the lattice has to be tilted in opposite directions
determined by $-\vec\zeta$. 
The BdG description of this system is,
\bea
\nn
H&=&\begin{pmatrix}
H_+ & \Delta \\
\Delta^\dagger & H_-
\end{pmatrix},
\label{bdg.eqn}
\eea
where
\bea
H_\pm &=& - i \hbar v_F (\sigma_x \partial_x \pm \sigma_y \partial_y) +U\\
& \mp&   i \hbar v_F \sigma_0(\zeta_x \partial_x + \zeta_y \partial_y ).
\label{H+-.eqn}
\eea
describes degrees of freedom near the valley labeled by $\tau=\pm$. Here 
$v_F$ is the isotropic Fermi velocity of excited carriers in non tilted graphene which in the tilted Dirac case is modified by
matrix proportional to $\sigma_0$ whose coefficient is given by the tilt velocity scale $\vec v_t\equiv v_F\vec \zeta $.  $U$ is an adjustable electrostatic potential that 
following Beenakker~\cite{Beenakker_SAR}, we choose it such that the Fermi wave vector in the normal region is much smaller than its value in superconductor and we adopt the step function profile for the pair potential $\Delta(\vec r)=\Delta_0 e^{i\Phi(\vec r)}$ at the interfaces of normal-superconductor junctions. 
The superconducting phase $\Phi(\vec r)$ is chosen as $-\phi/2$ ($\phi/2$) inside the left (right) superconducting lead. 
Eq.~\eqref{bdg.eqn} in normal and superconducting regions can be straightforwardly solved for both 
$\vec\zeta=0$~\cite{Beenakker_SAR} and arbitrary $\vec\zeta$~\cite{Faraei_PAR}. 
In the $\vec\zeta=0$ limit, the physics of AR becomes particularly transparent in two situations: 
(i) when the excitation energy ($\eps$) is much smaller than the chemical potential ($\mu$), 
one is dealing with an extended Fermi surface, and therefore the retro Andreev reflection is dominant. This defines the RAR-dominated regime.  
(ii) In the opposite regime, when $\eps \gg \mu$, the chemical potential $\mu$ will be the smaller energy scale of the problem. In this case the holes will tend to
arise from the lower part of the Dirac cone with opposite helicity and therefore the specular Andreev reflection prevails (Fig~\ref{AR.fig})~\cite{Beenakker_SAR}. 
In the $\vec\zeta\ne 0$ case, the generic effect of $\zeta$ is to bring the angle of the reflected hole closer to the normal in both $\eps\gg\mu$ and $\mu\gg\eps$ regimes.
In the extreme case of $\zeta\to 1$, the Andreev reflected hole tends to come very close to perpendicular to the interface direction for every value of 
incident electron angle~\cite{Faraei_PAR}.  

Regardless of whether we are in $\eps\gg\mu$ or $\mu\gg\eps$ regime, the condition for an electron with a specific energy $\eps$ and given $k_y$ 
to form an Andreev bound state corresponding to the semiclassical orbit starting from one of the interfaces and returning to the same point after
getting Andreev reflected at the other interface, is that the total phase change of the electron must be an integer multiple of $2\pi$. 
This quantization condition in our system is given by~\cite{Titov}:
\bea
\nn
&&
\bigg[ \cos(k_x^e d) \cos(k_x^h d) + \frac{\sin(k_x^e d) \sin(k_x^h d)}{\cos\alpha \cos\alpha'} \bigg] \cos (\frac{2\eps}{\Delta_0}), \\
\nn
&+&\bigg[ \frac{\sin(k_x^e d) \cos(k_x^h d)}{\cos\alpha} + \frac{\cos(k_x^e d) \sin(k_x^h d)}{\cos\alpha'} \bigg] \sin (\frac{2\eps}{\Delta_0}),\\
&+&\sin(k_x^e d) \cos(k_x^h d) \tan\alpha \tan\alpha'=\cos\phi,
\label{quantization.eqn}
\eea
where $k_x^{e(h)}$ is the $x$ component of the wave vector of the incident electron (reflected hole) and $\alpha^{(')}$ as depicted in Fig~\ref{model.fig} is the angle of the incidence (reflection). The angle $\alpha$ is given by $\tan^{-1}(k_y / k_x^e)$ and $\alpha'$ can be obtained from $\alpha$ by~\cite{Faraei_PAR}
\bea
\sin\alpha'&=&\frac{ (\eps+\zeta_x k_x)(\eps - \vec\zeta \cdot \vec k) }{ \zeta_x^2  k_x^2 + (\eps+\zeta_x k_x)^2}\label{aa'.eqn}\\
&\pm& \frac{ \zeta_x k_x \sqrt{\zeta_x^2 k_x^2 + (\eps+\zeta_x k_x)^2 - (\eps - \vec\zeta \cdot \vec k)^2 \sin\alpha^2} }{\zeta_x^2 k_x^2 + (\eps+\zeta_x k_x)^2}\nn.
\eea 
Note that the translational invariance along the borders implies $k_y$ to be the constant of motion, while $k_x^{e(h)}$ not only changes upon (Andreev) reflection,
but also explicitly depends on the valley index $\tau=\pm$ as follow,
\bea
[k_x^{e(h)}]^2 + k_y^2=[\eps\pm\mu \mp \tau(\zeta_x k_x^{e(h)}+\zeta_y k_y)]^2,
\label{kx_eq.eqn}
\eea
where the equation for hole is obtained from the corresponding equation of the electron by $(\mu,\tau)\to -(\mu,\tau)$.
When the above equation is viewed as a second order equation for the unknown $k_x^{e(h)}$, in the absence of tilt, i.e. with $\vec\zeta=0$, 
the solutions are symmetric under of $k_x \rightarrow -k_x$. This symmetry will be broken by turning on the tilt. 
We will return to the discussion of the consequences of this symmetry breaking in following section. 
Taking the valley attribute $\tau$ into account, Eq.~\eqref{kx_eq.eqn}, we obtain four solutions given by~\footnote{Note that in the $\vec\zeta=0$ limit,
it is not easy to discern such a four-fold degeneracy.},
\bea
&&k_x^e=  \label{kx.eqn}\\
&&\frac{-\tau \zeta_x (\eps+\mu - \tau \zeta_y k_y) \pm \sqrt{(\eps+\mu - \tau \zeta_y k_y)^2 - (1-\zeta_x^2) k_y^2} }{1-\zeta_x^2}.\nn
\eea
The above four solutions in the non-tilted case all have the same absolute value. 
For a hole of given energy and a fixed $k_y$, the corresponding solutions can be obtained by 
$\mu \rightarrow -\mu$ and $\tau \rightarrow -\tau$. 

As can be seen from Eq.~\eqref{kx.eqn}, the valley index and tilt parameters $\zeta_i$ appear together.
Therefore in the $\vec\zeta=0$ there will be no distinction between the solutions corresponding to the valleys $\tau=\pm$.
In this case, there will be two (two-fold degenerate) solutions related by $k_x\to -k_x$ symmetry. 
Because of this symmetry, in the SAR regime where Andreev modes can propagate along $y$ direction, 
electrons incident to the left and right interfaces contribute equally to the
Andreev mode which is equal to the contribution of the holes. Therefore no net charge current can be
carried by the Andreev modes. However, breaking the above symmetry by turning on $\vec\zeta$ 
makes the situation asymmetric between electrons and holes (note that the holes are obtained by
$\vec\zeta\to -\vec\zeta$). This electron-hole asymmetry leads to a net electric charge carried by Andreev modes 
along the $y$ direction in Fig.~\ref{model.fig}. In the following we analyze the 
quantization condition~\eqref{quantization.eqn}, in various limits to see how can the Andreev mode
become charged by tilting the Dirac cone.

\section{Long junction regime: Charged Andreev modes}
\label{ChargedAM.sec}
Let us start by considering the SAR regime. In the long junction regime where the superconducting coherence length is much smaller than the width of the normal 
channel ($\xi \ll d$), the Thouless energy, $E_T=\hbar v_F/d$ will satisfy $E_T \ll \Delta_0$. This will set the gap energy scale $\Delta_0$
as the larger of the relevant energy scales within which Andreev modes -- whose energy are comparable to Thouless energy scale -- can disperse. 
This allows us to simplify the quantization condition~\eqref{quantization.eqn} by assuming $\eps \ll \Delta_0$ to arrive at~\cite{Andreevmodesgap}:
\bea
\cos \phi &+& \cos(k_x^e d) \cos(k_x^h d) \label{simple_quantization.eqn} \\
&+& \bigg( \frac{1-\sin\alpha \sin\alpha'}{\cos \alpha\cos \alpha'} \bigg) \sin(k_x^e d) \sin(k_x^h d)=0. \nn
\eea
In the above expression, $k_x^e$ is given by Eq.~\eqref{kx.eqn} and the $k_x^h$ is the 
corresponding hole wave vector obtained by $(\tau,\mu)\to -(\tau,\mu)$ in Eq.~\eqref{kx.eqn}.
This is equivalent to replacing $\eps$ by $-\eps$~\cite{Titov}.

\begin{figure}[t]
\centering
\includegraphics[width=8cm]{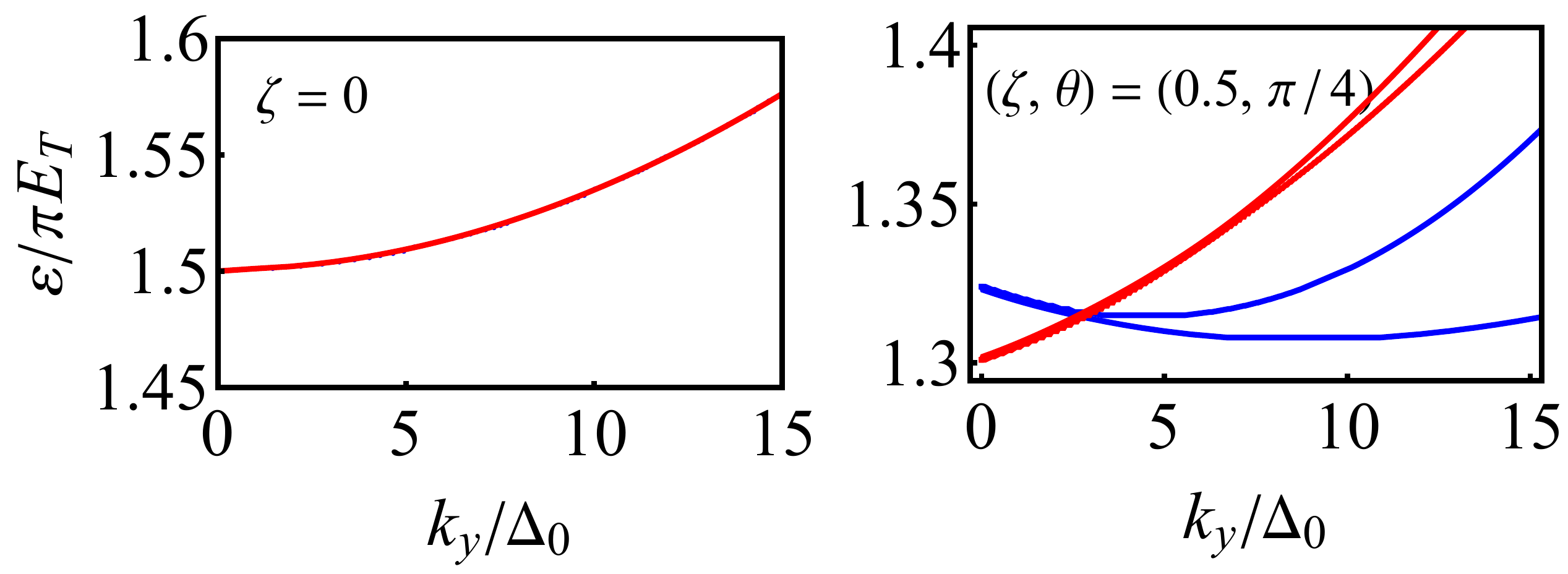}
\caption{Dispersion of Andreev modes. Energy of Andreev modes are naturally measured in the units of $\pi E_T$ and momenta are measured in units 
of $\Delta_0/(\hbar v_F)$ where the superconducting gap parameter $\Delta_0$ is set as a global unit of energy. The chemical potential $\mu=0.1\Delta_0$,
the phase difference $\phi$ of the two superconductors is assumed to be zero. 
Left panel corresponds to $\zeta=0$ while the right panel corresponds to $\zeta=0.5$ and $\theta=\pi/4$. The tilt $\zeta$ clearly 
breaks the four-fold degeneracy of the left panel. The blue and red curves correspond to the sings $\pm$ proceeding the square root in Eq.~\eqref{kx.eqn}.
The bifurcation-like feature within each color arises from the valley index $\tau$. 
}
\label{e_ky.fig}
\end{figure}

\subsection{Splitting of four-fold degeneracy of Andreev modes}
For a given $k_y$, the above quantization condition admits multitude of solutions $\eps_n$ each labeled by a branch (Andreev band) label $n$. 
For a given branch labeled by $n$, in the untilted case ($\zeta=0$), the four solutions (a factor of $2$ comes from valley index $\tau$, and another
factor of $2$ comes from $\pm$ labeling the two solutions of the quadratic equation~\eqref{kx_eq.eqn}) for $k_x^{e(h)}$ in Eq.~\eqref{kx.eqn} will have the same absolute values
which are related by sign reversal of horizontal momentum. But the above quantization condition 
does not care about the sign reversal of the solutions $k_x^e$ and $k_x^h$ of Eq.~\eqref{kx.eqn}. 
Therefore in the $\vec\zeta=0$ limit, each branch will be fourfold degenerate. 
This can be clearly seen for the typical ($n=2$) branch in Fig.~\ref{e_ky.fig}. As can be seen in the left panel, all four
solutions coincide. The degeneracy is lifted upon turning on a $\zeta=0.5$ tilt along the $x$-axis (i.e. $\theta=0$). 

The distinct Andreev mode bands are denoted by two sets of blue and red color in the right panel of Fig.~\ref{e_ky.fig}.
Further splitting within each color which looks like a "bifurcation"~\footnote{Note that, technically speaking, the bifurcation is a feature of nonlinear systems~\cite{Strogatz},
and is not necessarily related to degeneracy of linear operators. However, to emphasize this particular splitting, let us use the word bifurcation to denote
this portion of degeneracy lifting.} arises from the valley index $\tau$ in Eq.~\eqref{kx.eqn}. 
The blue (red) correspond to $+$ ($-$) sign in front of the square root in Eq.~\eqref{kx.eqn}. 
This square root in $\zeta\ne 0$ case is responsible for breaking the $k_x\to -k_x$ symmetry. 
From now on, we refer to this $\pm$ sign as the color attribute. Later on, we will show that
as far as propagating Andreev modes are concerned, only a portion of red (blue) modes will be
relevant that satisfy $k_x<0$ ($k_x>0$).

In the $\vec\zeta=0$ case, the colors further signify whether the electron is incident upon the right superconductor (blue)
or on the left superconductor (red). Since Eq.~\eqref{simple_quantization.eqn} is insensitive to the color sign (of $k_x^{e(h)}$), 
the right-moving (blue) and left-moving (red) electrons will experience quite symmetric situations (arising from $k_x\to -k_x$ symmetry) 
which will therefore result in the four-fold degeneracy in the left panel of Fig.~\ref{e_ky.fig}. 
Upon turning on the tilt, as a result of breaking the $k_x\to -k_x$, the right and left-movers will not experience symmetric conditions anymore. 
In presence of $\vec\zeta$, being a right- or left-mover depends on the $\vec\zeta$. 
Indeed it is clear from Eq.~\eqref{kx.eqn} that the $x$-component $\zeta_x$ of tilt vector $\vec\zeta$ plays the essential role in 
breaking the symmetry of  $k_x \rightarrow - k_x$. 

\begin{figure}[t]
\centering
\includegraphics[width=8.5cm]{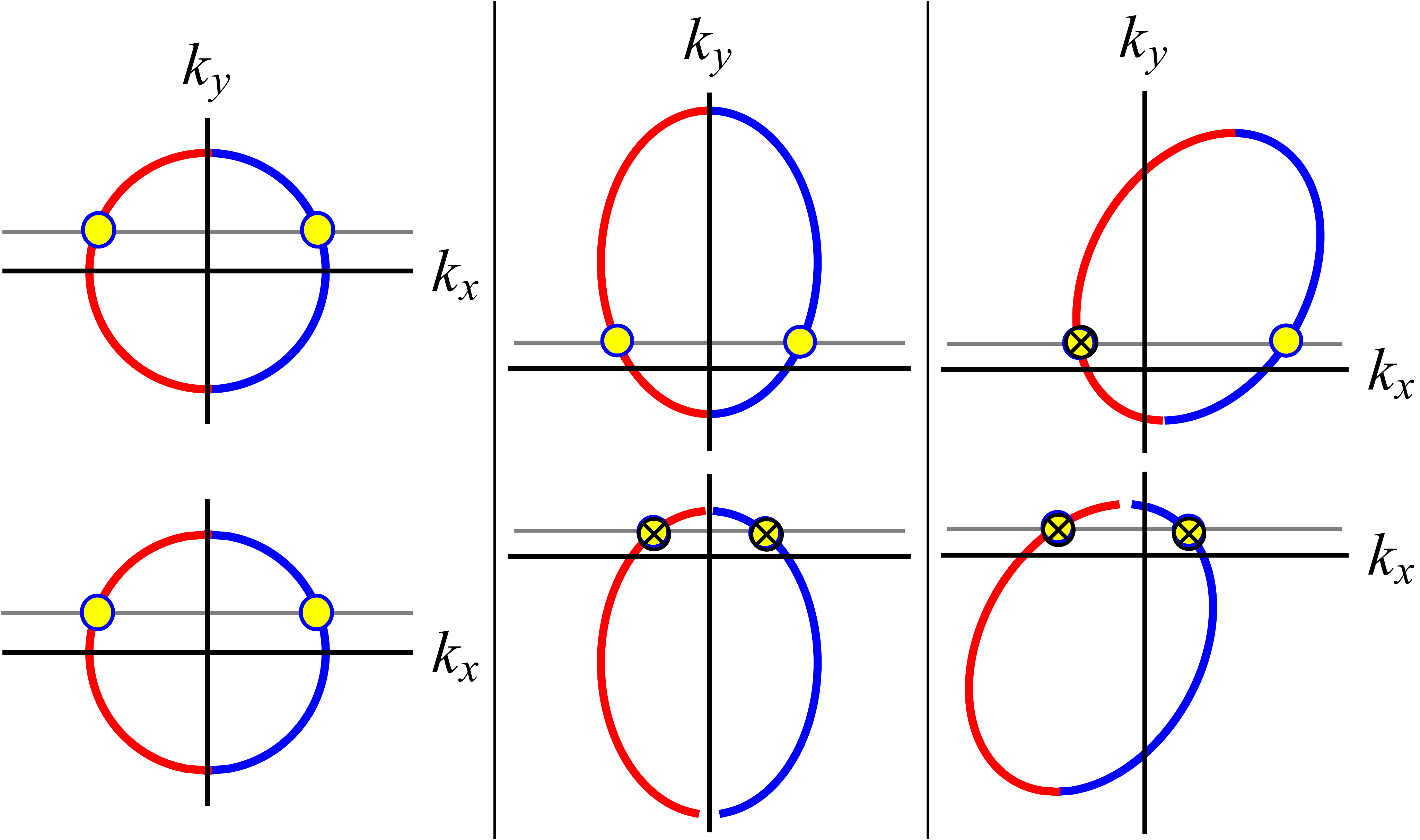}
\caption{Constant energy contours in $k_xk_y$ space before imposing the quantization condition. Top and bottom energy contours
correspond to the two valleys (which are separated along $y$-axis in our model). The left, middle and right panels correspond to
$(\zeta_x,\zeta_y)=(0,0)$, $(0,0.7)$ and $0.7(1/2,\sqrt 3/2)$, respectively. The gray (horizontal) line denotes a constant $k_y$. 
Upon quantization, those $k_x$ solutions that satisfy (do not satisfy) the quantization condition~\eqref{quantization.eqn} are denoted by small open
circle (cross).}
\label{cones.fig}
\end{figure}

\begin{figure}[t]
\centering
\includegraphics[width=6cm]{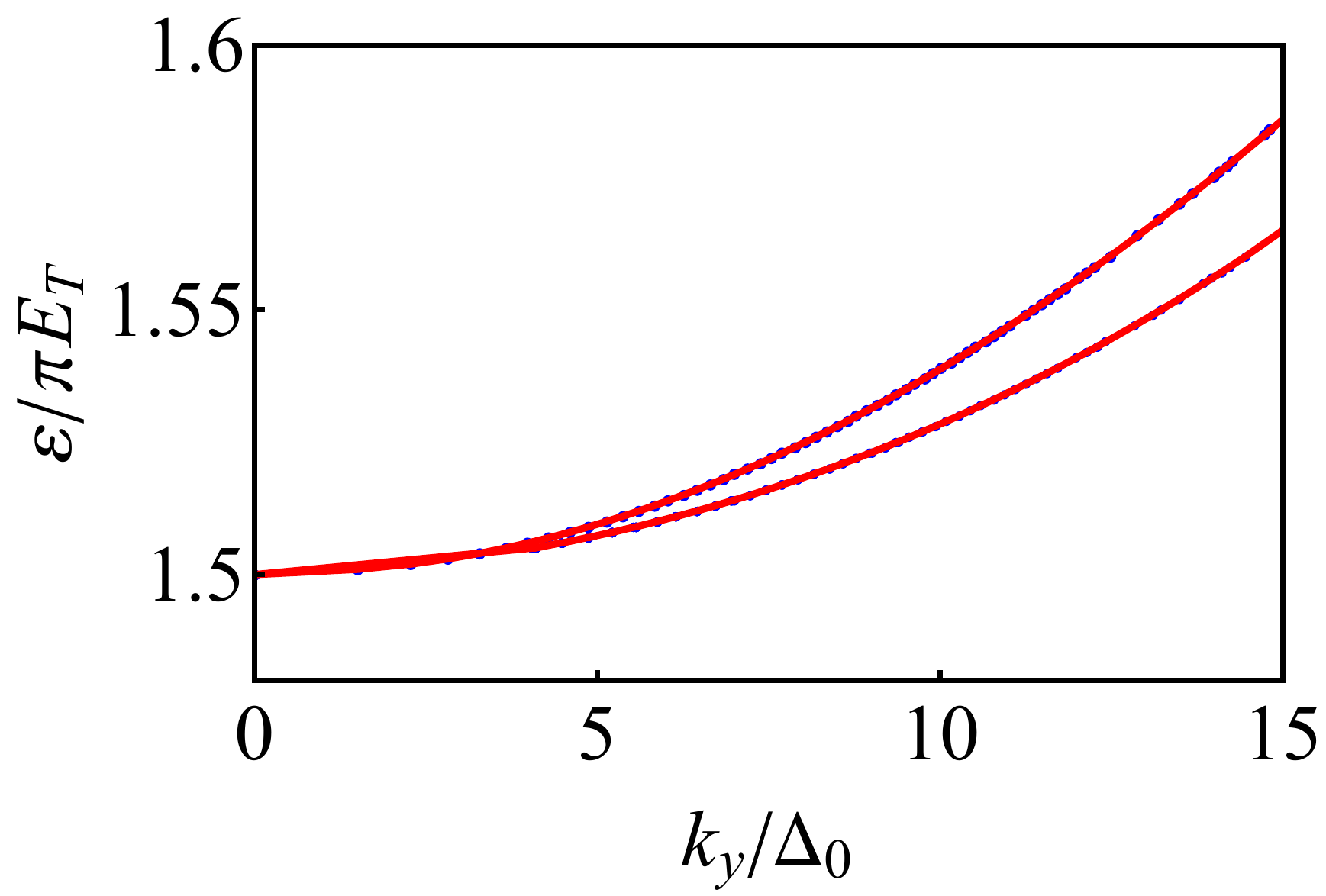}
\caption{For a tilt along $y$ direction, since the $k_x\to -k_x$ symmetry persists, the color degeneracy
will persist. The "bifurcation"-like splitting arises from the asymmetry between the cones (breaking of $k_y\to -k_y$ symmetry). 
This confirms that the bifurcation-like splitting is due to the asymmetry between the valleys. }
\label{bifurcation.fig}
\end{figure}

In Fig.~\ref{cones.fig}, we have schematically represented the constant energy contours for the two colors ($\pm$) before imposing
the quantization of the $k_x$ component of momentum. In the left panel tilt is zero. In the middle panel, the tilt is along
$k_y$ direction, and therefore still the $k_x\to -k_x$ symmetry is intact. In the right panel the tilt is along a generic
direction and breaks the $k_x\to -k_x$ symmetry. Therefore for a given $k_y$ value (gray horizontal line), if the $k_x$ solution
at the interface of blue and gray line satisfies the quantization condition~\eqref{quantization.eqn}, the red one is not obliged
to satisfy it. Those intersections satisfying Eq.~\eqref{quantization.eqn} are denoted by open circles. 
That is why in the left panel which is still symmetric under $k_x\to -k_x$, all four intersections satisfy Eq.~\eqref{quantization.eqn}
and are therefore denoted by open circles. Let us move now to the middle panel where $\zeta_x=0,\zeta_y=0.7$.
It can be guessed and it is indeed true that vertical tilts which are in $k_y$ direction do not change this picture.
However, breaking the $k_y \to -k_y$ means that if the two crossings at the top valley are solutions of Eq.~\eqref{quantization.eqn}, 
those in the bottom panel will not satisfy it. 
In this situation, regardless of the existence of the tilt, the blue and red electrons give rise to degenerate modes. 
Therefore in this case the splitting between the blue and red colors disappears. The other valley will contribute a mode
at a different energy. Therefore in the middle panel, the valley splitting persists. To confirm this, in Fig.~\ref{bifurcation.fig}, 
we have plotted the solutions of Andreev mode, Eq.~\eqref{quantization.eqn} for a tilt along the direction $y$ along which the two 
tilted Dirac cones are separated in the Brillouin zone. As can be seen in Fig.~\ref{bifurcation.fig}, 
the colors are degenerate, while the bifurcation-like feature related to the valley index survives. 
Finally let us return to the right panel of Fig.~\ref{cones.fig} where the tilt parameter is a generic vector
given by $\vec\zeta=0.7(1/2,\sqrt 3/2)$. In this case both $k_x\to -k_x$ and $k_y\to -k_y$ symmetries are broken,
and therefore out of the four degenerate solutions (four open circles) of the left panel, only one will satisfy 
the quantization condition~\eqref{quantization.eqn}. This explains the origin of color splitting and valley bifurcation
in Fig.~\ref{e_ky.fig}.

The band edges for the Andreev modes can be analytically calculated. To calculate them one simply
needs to evaluate the energies $\eps_n$ corresponding to $k_y=0$. In this limit, since the conserved 
$k_y$ is zero, it follows that $\alpha=0$ and $\alpha'=\pi$, and therefore the quantization relation~\eqref{simple_quantization.eqn}
reduces to $\cos\phi+\cos(k_x^e+k_x^h)d=0$. But Eq.~\eqref{kx.eqn} at $k_y=0$ reduces to
$$
   k_x^{e}(\eps)=\frac{(\eps+\mu)(\lambda-\zeta_x)}{1-\zeta_x^2},
$$
where $\lambda$ is the color index. Since $k_x^h=k_x^e(-\eps)$, the quantization condition for $k_y=0$ gives the following sets of discrete energies $\eps_n$ satisfying
$$ 
  \cos\left[\frac{2( \eps_n- \lambda \mu \zeta_x) d}{1-\zeta_x^2 }\right]+\cos\phi=0
$$
which readily gives the following solutions:
\be
   \frac{2(\eps_n-\lambda \mu \zeta_x) d}{1-\zeta_x^2 }=(2n+1)\pi-|\phi|,~~~ n=0,1,\ldots,
\ee
where the principal value of the phase difference $|\phi|$ is defined to be between $0$ and $\phi$. 
Upon restoring constants $\hbar$ and $v_F$, the above
energies will be naturally expressed in units of the Thouless energy $E_T$.  
Therefore we obtain,
\be
   \eps_n=\lambda \mu \zeta_x + \frac{1-\zeta_x^2}{2}\left[(2n+1)\pi- \phi \right] E_T.
   \label{bandedge.eqn}
\ee
This equation is a nicely generalization Eq.~(2) of Ref.~\cite{Andreevmodesgap} in two respects:
(i) instead of $n=0$ (lowest mode), it is valid for arbitrary $n$, and (ii) it includes the effects of
tilt parameter $\vec\zeta$ and reduces to the result of Ref.~\cite{Andreevmodesgap} by setting $\vec\zeta=0$.
For arbitrary $n$ this equation gives the band edges of the Andreev modes in tilted Dirac fermions and the 
splitting arising from the tilt $\zeta_x$ is naturally encoded into the above equation. 
Note how the color index, $\lambda=\pm1$ naturally appears in this equation. The above equation is in agreement
with Fig.~\ref{e_ky.fig}. Furthermore, when $\zeta_x=0$, in agreement with Fig.~\ref{bifurcation.fig}, the
color index becomes irrelevant and blue and red colors coincide. 

The Andreev modes (e.g. in Fig.~\ref{e_ky.fig}) starting at $\eps_n$, disperse linearly around $k_y=0$.
To analytically calculate the slope (velocity) associated with this linear dispersion, one needs to 
repeat the above procedure up to first order in $k_y$. Eq.~\eqref{kx.eqn} up to this order gives,
\bea
k_x^e(\pm\epsilon)=\frac{ \epsilon\pm \mu \mp \zeta_y k_y}{\lambda \pm \zeta_x}. 
\eea
Using the above value in quantization Eq.~\eqref{quantization.eqn} gives,
\bea
   v_\zeta = -\lambda\tau \zeta_x \zeta_y. 
   \label{vy.eqn}
\eea
For example in Fig.~\ref{bifurcation.fig} where $\zeta_x=0$, one can nicely see that the slope is
indeed zero. The velocity~\eqref{vy.eqn} of Andreev modes around the $k_y=0$ will approximately
replace the average velocity $\bar v$ of Ref.~\cite{Andreevmodesgap} by $\bar v-\lambda\tau\zeta_x\zeta_y/2$. 

To summarize, by breaking $k_x\to -k_x$ the colors split, and by breaking $k_y\to -k_y$ the bifurcation-like
splitting related to the valley index $\tau$ appears. The later splitting exists for any-type of tilt, while the former
splitting requires a non-zero $\zeta_x$. In a generic situation where a vector $\vec b$ connects the two tilted Dirac cones around the two valleys,
the component of $\vec\zeta$ which is longitudinal to $\vec b$ can only generate the bifurcation-like (valley) splitting, leaving 
the color degeneracy intact. The transverse component will split both colors and valleys. 

\begin{figure}[t]
\centering
\includegraphics[width=8cm]{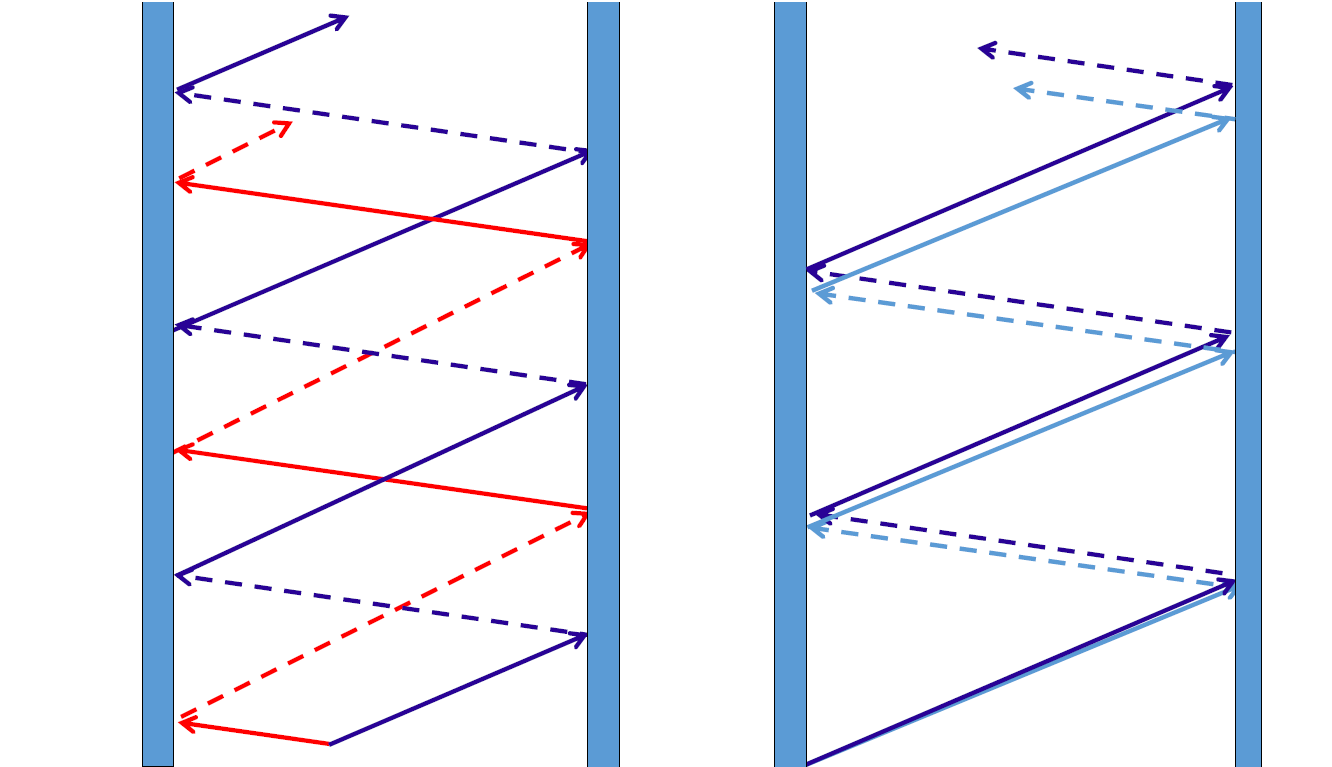}
\caption{Calculated semiclassical trajectories of Andreev modes for a fixed $k_y$ momentum along the channel. 
Blue and red colors have the same meaning as in Fig.~\ref{e_ky.fig}. Solid (dashed) lines correspond to electrons (holes). 
(Left) Both modes belong to the same valley (top valley, $\tau=+1$).
(Right) For the same (blue) color corresponding to electrons incident on the right interface, two arrows 
correspond to two different valleys. Dark (light) blue corresponds to $\tau=+1$ ($-1$). These two semi-classical 
paths correspond to the bifurcation-like splitting. }
\label{blured.fig}
\end{figure}

\subsection{Semiclassical trajectories of Andreev modes}
Due to translational invariance along $y$ direction, $k_y$ is a constant of motion. 
In this sub-section we would like to study the splitting of semiclassical trajectories of Andreev modes
upon turning on the tilt parameter $\vec\zeta$. For a given $k_y$, there are four degenerate modes
which will split upon turning on the tilt $\vec\zeta$ as was demonstrated in Fig.~\ref{e_ky.fig}.  
For a fixed $k_y$, there are four electrons (of course with four different energies) that satisfy the quantization condition~\eqref{quantization.eqn};
two right movers with two different $\tau$ (valley) indices, and two left-movers from each valley. 
Having fixed the conserved quantum number $k_y$, the solution $k_x$ will determine the incident angle of the electrons. 
In the left panel of Fig.~\ref{blured.fig}, we have plotted the calculated angles for two of the above solutions 
corresponding to valley index $\tau=+1$ (solid lines). There are two more solutions corresponding to $\tau=-1$ which is not shown
in the left panel. The quantized $k_x$ for holes is similarly calculated from Eq.~\eqref{kx_eq.eqn}. The resulting 
angles give rise to dashed lines in the left panel of Fig.~\ref{blured.fig}. Solid and dashed lines carry
opposite charges. This picture is valid for $k_y>0$. A similar picture
can be constructed for $k_y<0$. In the absence of thermal or electro-chemical gradient, both 
$k_y>0$ (upward moving) and $k_y<0$ (downward moving) modes will have equal chances.
But in the presence of a such gradients along $y$ direction, the upward moving modes 
shown in the left panel of Fig.~\ref{blured.fig} will be dominate. 
Now, when $\vec\zeta$ is zero, this figure will have a left-right symmetry. Therefore the net charge 
carried by such a mode (charge of dotted lines minus charge of solid lines) will be zero. Upon tilting the Dirac cone, the entire semiclassical paths will be tilted
in such a way to generate the left panel of Fig.~\ref{blured.fig}. 
This is how, a charge imbalance in the Andreev mode
is generated. Therefore, the putatively charge neutral Andreev modes acquire a charge upon tilting the Dirac cone.
So far, in the left panel, we have only considered the current from one ($\tau=+1$) valley. 
To investigate the role of other valley, 
in the left panel of Fig.~\ref{blured.fig}, we have plotted two blue solutions. Dark (light) blue correspond to $\tau=+1$ ($\tau=-1$) valley. 
For clarity, in this panel we have ignored the red (left moving) solutions. This panel shows the real space manifestation of 
the bifurcation-like splitting that stems from valley degeneracies. This panel clearly shows that the contribution of the other
valley to the net current is additive. 

The above intuitive picture for the charge current of Andreev modes can now be put on a formal
setting in the next sub-section. 


\subsection{Calculation of charge current by Andreev mode}
To formally see how a non-zero electric current can arise from the tilt parameter and vanishes by 
setting $\vec\zeta=0$, let us argue in terms of a Landauer-B\"uttiker formulation.
The electrical current due to propagating Andreev modes incident at angle $\alpha$ is determined by two essential factors,
namely the velocity matrix element $v_y(\alpha)$ 
of Andreev modes 
and the density $g(\alpha)=\frac{\cos\alpha + \zeta\cos\theta}{[1+\zeta\cos(\theta+\alpha)]^2}$ of modes as 
$\langle\langle J_y\rangle\rangle= e g(\alpha) \langle v_y(\alpha)\rangle$~\cite{Faraei_PAR}
where $e$ is the charge of electron. The total current is integral of the above expression over the 
permissible range of angles $\alpha$. 
The velocity matrix element due to a mode incident at an angle $\alpha$ is $j_y(\alpha)=e\langle v_y(\alpha)\rangle$ 
where the $\langle ... \rangle$ indicates the quantum average with respect to the scattering states of the tilted-Dirac-BdG equation,
\bea
j_y(\alpha)=2 e v_F \big[ \frac{\zeta_y}{\cos\alpha}+\frac{\zeta_y}{\cos\alpha'}+(\tan\alpha-\tan\alpha') \big].
\label{j.eqn}
\eea
Here $\alpha'$ is the angle of the Andreev reflected hole which in the SAR regime is given by Eq.~\eqref{aa'.eqn}. 
In the above expressions there are two distinct contributions to the charge current. The first term, directly 
depends on $\zeta_y$, while the second term stems from the difference in the incident angle $\alpha$ of electron
and the reflection angle $\alpha'$ of the hole. When there is not tilt, these two angles are equal. Therefore the 
difference in the angles will be an indirect contribution of the tilt parameter in the charge current carried by
Andreev modes. This expression formally shows how an electric current can arise from the tilt which
would otherwise vanish. 

Let us now turn into a simple Boltzmann kinetic treatment~\cite{GirvinBook}. 
In the kinetic description, 
to proceed with the calculation of transport coefficients, we need to integrate 
appropriate moments of the above group velocity weighted by appropriate power of energy, 
over the (fermionic) Andreev bands. For ballistic transport, the transport coefficients are determined by~\cite{MarderBook}
$$\mathfrak{L}_{ij}^{(\nu)} = e^2 \int dk_y \frac{\pr f}{\pr \mu} v_i v_j \left[\eps^\nu(k_y)-\mu_a\right] $$
In this semiclassical equation $\eps(k_y)$ is the dispersion of Andreev modes (see e.g. Fig.~\ref{e_ky.fig} or Fig.~\ref{bifurcation.fig}) 
from which the group velocities are derived as $v_i=\pr \eps/ \pr k_i$ with $i=x,y$, and $f$ is the equilibrium Fermi-Dirac occupation probability. 
To emphasize that this is a transport theory for Andreev modes, we have explicitly included the $\mu_a=0$ of the fermions corresponding to Andreev modes. 
The transport coefficients obtained from the above equations relate the gradient of electrochemical potential and temperature
to charge and heat currents as~\cite{GirvinBook}
\be
   \begin{pmatrix}
      \vec j \\ \vec j_Q
   \end{pmatrix}
   =
   \begin{pmatrix}
   {\mathfrak L}^{(0)} 	& -e^{-1}{\mathfrak L}^{(1)}   \\
   -e^{-1}{\mathfrak L}^{(1)} 	& e^{-2} {\mathfrak L}^{(2)}  
   \end{pmatrix}
   \begin{pmatrix}
      \vec \Sigma  \\ -\frac{\vec\nabla T}{T}
   \end{pmatrix}
   \label{thermoelectric.eqn}
\ee
where $\vec\Sigma=\vec E-e\vec\nabla \mu$. 
For a generic dispersion $\eps(k_y)$ of Andreev modes, an average $k_y$-independent velocity $\bar v$ was 
employed in Ref.~\cite{Andreevmodesgap} to calculate the ${\mathfrak L}^{(2)}$. The rest of ${\mathfrak L}^{\nu}$s with
$\nu=0,1$ for upright Dirac fermions become zero. This is due to charge neutrality of Andreev modes in upright Dirac cones. 
Let us now consider corrections to the transport coefficients of the upright Dirac cones. 
In order to do so, at low temperatures one can focus on the low $k_y$ part of the dispersion of $n=0,1$ branches of the Andreev modes in tilted Dirac cones. 
As pointed out below Eq.~\eqref{vy.eqn}, the transport coefficients obtained in Ref.~\cite{Andreevmodesgap} are 
corrected by tilt term via $\bar v\to \bar v+v_\zeta/2$. Assuming that the first branch dominates the transport, 
in terms of the energy $\eps_0^{\zeta=0}=(\pi-\phi) E_T$ of 
the first branch of the upright Dirac cone~\cite{Andreevmodesgap} and summing over the two colors $\lambda=\pm 1$ of the first branch we obtain,
\bea
&& \delta{\mathfrak L}^{(0)}= e^2 \zeta_x^2 \zeta_y  \mu \label{L0.eqn}\\
&&\delta {\mathfrak L}^{(1)}=- e \zeta_x^2 \zeta_y \frac{1-\zeta_x^2}{2}\eps_0^{\zeta=0} \label{L1.eqn}\\
&&\delta {\mathfrak L}^{(2)}= \zeta_x^2 \zeta_y \mu \left[\frac{\mu^2\zeta_x^2}{3}+\frac{(1-\zeta_x^2)^2}{4}\left(\eps_0^{\zeta=0}\right)^2 \right]\label{L2.eqn}
\eea
Note that, due to charge neutrality of the Andreev modes, the values of ${\mathfrak L}^{(0,1)}$ for an upright Dirac cone that
arises from Andreev modes is already zero. The charge conductance in Eq.~\eqref{L0.eqn} is the tilt-induced correction to the
charge conductance. But since in addition to Andreev modes, the Bloch electrons also contribute to the charge conductance,
and that it does not depend on the phase difference $\phi$ of the two superconductors, separation of this effect from
other background electric currents in a real experimental situation can be challenging. 
Eq.~\eqref{L2.eqn} provides a correction to the thermal conductance of upright Dirac cone calculated in Ref.~\cite{Andreevmodesgap}~\footnote{
Note that the thermal conductance $\kappa$ defined by $\vec j_Q=\kappa (-\vec\nabla T)$ under the condition of no electric current flow, implies that
$e^2\kappa={\mathfrak L}^{(2)}-\left({\mathfrak L}^{(1)}\right)^2/{\mathfrak L}^{(0)}$ in Fermi liquids where due to a finite $\mu$ one has (Sommerfeld expansion) 
${\mathfrak L}^{(1)}\propto T^2$, at low enough temperatures will be simply given by ${\mathfrak L}^{(2)}$~\cite{MarderBook,GirvinBook}. However, in the
present case where the chemical potential of Andreev modes is zero, one has to use this full expression to compare the tilt-induced measurements. 
}.
The dependence of this term on the phase difference $\phi$ of superconductors (encoded into $\eps_0^{\zeta=0}$)
distinguishes them from contribution of normal electrons. 

The most striking term comes from Eq.~\eqref{L1.eqn}. This equation provides a correction to a charge conductance
obtained from thermal gradient. This quantity is zero for upright Dirac cone. As such, this correction is the largest 
effect. This term describes the transport of charge by Andreev modes that would have been otherwise charge neutral. 
The entire effect comes from the tilt. The dependence of the charge conductance on the phase difference $\phi$ of superconductors
encoded into $\eps_0^{\zeta=0}$ is the factor that distinguishes this contribution from the charge transport caused by 
normal electrons in response to temperature gradient.

In the passing, let us observe an interesting property of the transport coefficients of 
tilted Dirac fermions in SNS setup. The above corrections satisfy 
$\delta{\mathfrak L}^{(\nu)}(\vec\zeta)=-\delta{\mathfrak L}^{(\nu)}(-\vec\zeta)$. 
This is reminiscent of the Onsager reciprocity relation 
${\mathfrak L}^{(\nu)}(\vec B)=-{\mathfrak L}^{(\nu)}(-\vec B)$. 
To make this sound plausible, let us note that a crossed $\vec E$ and $\vec B$ configuration gives 
rise to a semiclassical drift velocity proportional to $\vec E \times \vec B$~\cite{GirvinBook}. Therefore
the Onsager reciprocity relation can be reinterpreted as odd dependence of the transport coefficients
on the semi-classical drift velocity. Now it remains to establish a connection between the drift velocity 
and the parameter $\vec \zeta$. This can be most easily seen if one notes that the energy-momentum dispersion 
relation in tilted Dirac equation can be written in terms of the metric 
$ds^2=-dt^2+(d\vec r- v_F\vec \zeta dt)^2$~\cite{JafariLorentz,Saharmetric,SaharPolariton,tohid,Volovikblackhole}.
This metric is equivalent to applying a Galilean boost $d\vec r\to d\vec r-v_F\vec\zeta dt$ to the Minkowski metric $ds^2=-dt^2+d\vec r^2$.
Therefore $v_F\vec \zeta$ can be interpreted as a drift velocity~\footnote{In Ref.~\cite{Rostamzadeh}, the $\vec\zeta$ is attributed to 
some sort of incipient electric field.}

\section{Short junction regime: Infinite density of states}
\label{InfDOS.sec}
Another relevant regime of SNS junctions is the short junction regime where $d \ll \xi $.
Let us investigate this regime in our setup based on the tilted Dirac electrons. 
In the short junction regime even when $\eps>\mu$ the system does not sustain propagative Andreev modes~\cite{Andreevmodesgap}. 
To see how the picture of propagating Andreev modes ceases to hold in the short junction limit, one can utilize
Eq.~\eqref{bandedge.eqn}, according to which the energy scale of propagating Andreev modes are set by the 
Thouless energy $E_T$. In the short junction limit, $E_T$ exceeds the superconducting gap scale $\Delta_0$, and therefore
the putative Andreev modes will become part of the continuum of Bogoliubov excitations above the superconducting gap. 
In the short junction regime, sub-gap Andreev excitations will not propagate anymore, but they re-organized
themselves into localized Andreev levels. 
In this regime the localized Andreev levels correspond to closed semiclassical paths. Every such path involves 
an electron incident in an SN interface. The (retro) Andreev reflected hole will travel up to the other SN interface where
it undergoes another (retro) Andreev reflection and is reflected as an electron as in left panel of Fig.~\ref{AR.fig}. 

The same quantization condition Eq.~\eqref{quantization.eqn} is valid for these Andreev levels, as well. 
To solve the quantization condition, one notes that since a large energy scale $E_T$ governs the formation of Andreev levels at energy $\eps$, 
one can approximate $k_x^{e(h)}(\eps)$ by $k_x^{e(h)}(0)$, and $\alpha^{(')}(\eps)$ by $\alpha^{(')}(0)$~\cite{Titov}. 
The subsequent terms of the expansion are of the order of $(\Delta_0/E_T)^n$ with $n\ge 1$ which are negligible in short junction regime. This approximation simplifies the quantization condition to:
\bea
\eps =\Delta_0 \sqrt{ 1 + \frac{\mathfrak t}{2} (\cos \phi -1)}.
\eea
This equation is identical to the case of upright Dirac fermions in graphene~\cite{Titov}.
The transmission coefficient $\mathfrak t$ contains all the effects of tilt, $\vec \zeta$ and is given by~\cite{Titov},
\bea
{\mathfrak t}=\frac{{k_x^e}^2}{{k_x^e}^2 \cos^2(k_x^e d)  +(\tilde{\mu}/\hbar v_F)^2 \sin^2(k_x^e d)}.
\label{t.eqn}
\eea
Again the functional form of this expression is the same as upright Dirac cone~\cite{Titov} with the difference that 
$\mu \rightarrow \tilde{\mu}=\mu-\vec\zeta \cdot \vec k$. Furthermore the kinematics includes the effect of tilt and gives rise to the $k _x^e$ given by Eq.~\eqref{kx.eqn}.  Also the maximum value of the incident angle $\alpha$ of the electron contains the effect of $\vec\zeta$.
As for $k_x^e$, we have discussed its detailed dependence on the tilt parameter. But for the maximum value of $\alpha$, we recall that the 
only valid values for $\alpha$ are those that correspond to real values of $\alpha'$ in Eq.~\eqref{aa'.eqn}. 
For complex values of $\alpha'$, upon each AR, the amplitude of the initially
incident particle will be exponentially suppressed. Therefore after successive AR processes the amplitude of the Andreev bound state vanishes
which prevents the formation of Andreev bound states.
This condition will limit the values of $\alpha$ or equivalently the values of $k_y$ according to:
\bea
 \bigg( \frac{\zeta_y + \sqrt{1-\zeta_x^2 }}{\zeta^2-1}\bigg) \mu 
 \le  k_y \le  
 \bigg( \frac{\zeta_y - \sqrt{1-\zeta_x^2}}{\zeta^2-1}\bigg) \mu.
 \label{qeid.eqn}
\eea
\begin{figure}[t]
\centering
\includegraphics[width=7cm]{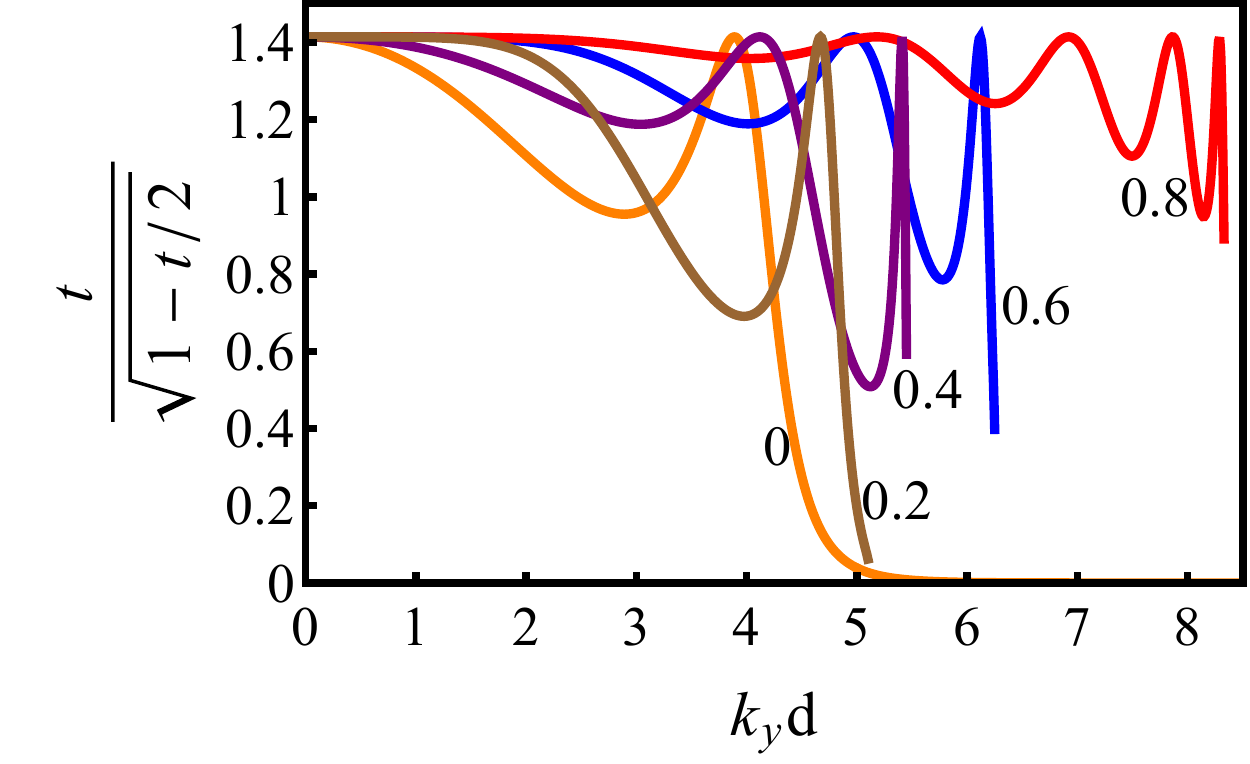}
\caption{The summand of the Josephson current at $\mu=-5 E_T$ for $\theta=0$ and $\zeta=0,0.2,0.4,0.6,0.8$. The limitations of the values of $k_y$ follow Eq.~\eqref{qeid.eqn}}
\label{sumand.fig}
\end{figure}
Note that we are working in units where $\hbar=v_F=1$.

\begin{figure}[t]
\centering
\includegraphics[width=7cm]{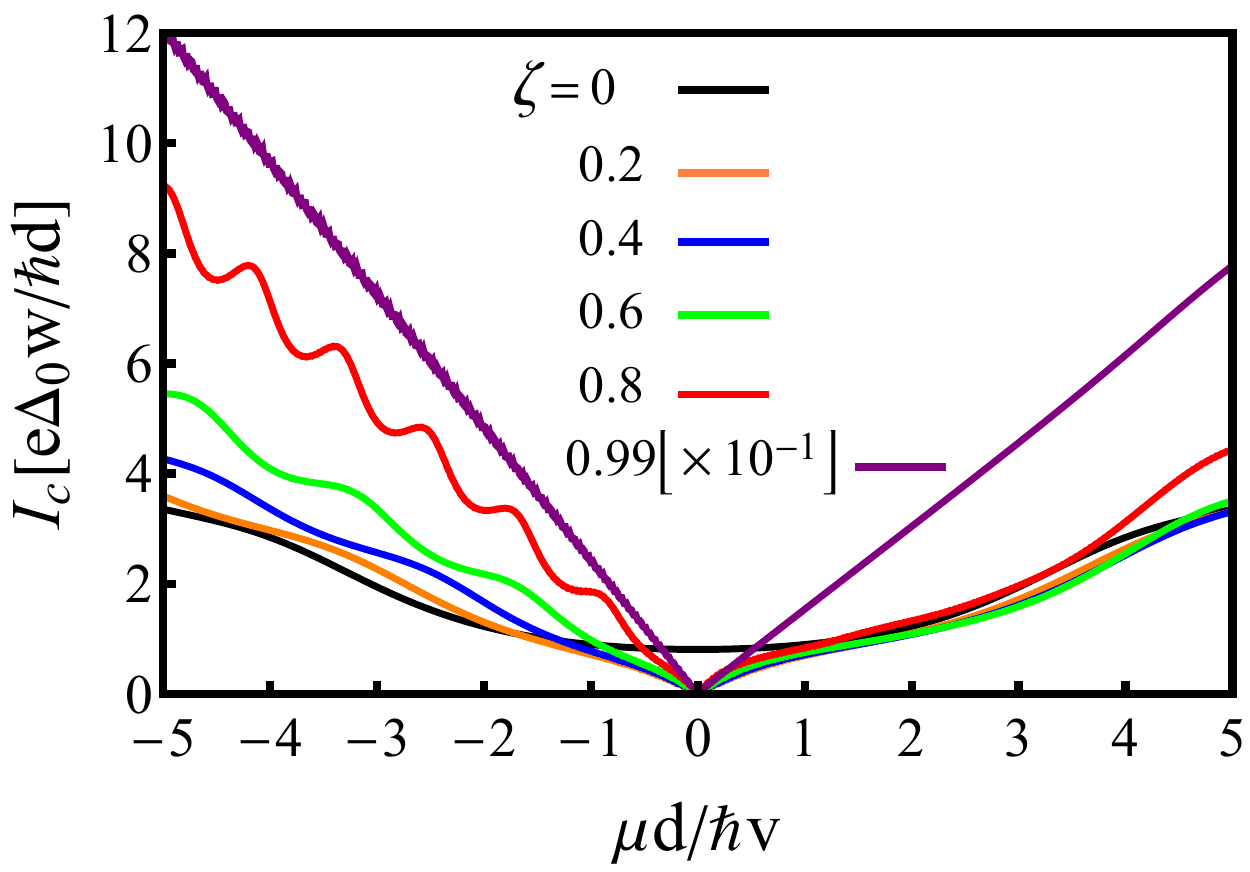}
\caption{Behavior of the critical Josephson current as a function of gate tunable chemical potential $\mu$ for various
values of the tilt parameter $\zeta$ indicated in the legend. The direction of the tilt is given by $\theta=\pi/4$.
The curve corresponding to $\zeta=0.99$ has been divided by $10$ to fit in the scale of the other plots. }
\label{Imu.fig}
\end{figure}

Now, we are ready to calculate the Josephson current and its dependence on the tilt parameter $\vec \zeta$. The current is which is given by~\cite{Titov}
\bea
 I= e\Delta \sum_n \frac{{\mathfrak t}\sin\phi}{(1- {\mathfrak t}/2 \sin\phi/2)^{1/2}},
 \label{I.eqn}
\eea
where $n$ labels the discrete Andreev bound states. Fig.~\ref{sumand.fig} shows the dependence of the summand in Eq.~\eqref{I.eqn} on $k_yd$ 
for values of the tilt parameter indicated in the figure. For tilt values not so close to $1$, the constraint~\eqref{qeid.eqn}
causes the plots to cease at some upper limit given in this equation. The integration of the summand in this case will 
rapidly converge. The resulting integration has been shown in Fig~\ref{Imu.fig}. 
As can be seen, upon increasing the tilt parameter $\zeta$,
the critical Josephson current significantly increases.  By approaching the $\zeta=1$ limit, the dispersion relation 
of tilted Dirac fermions will develop a flat band in the tilt direction. This gives rise to an enhancement of the density of states.
The question is, to what extent the enhancement of Josephson current in Fig~\ref{Imu.fig} is related to the enhancement of DOS due to band flattening.

To investigate this, let us focus on the extreme case of $\zeta=1$. In this situation, the Eq.~\eqref{qeid.eqn} will not impose any constraint on the allowed
$k_y$ values. This trend can be observed in Fig.~\ref{sumand.fig} where the support of the summand expands by bringing $\zeta$ closer to $1$. 
Hence in $\zeta=1$ situation where the constraint~\eqref{qeid.eqn} becomes inert, 
the behavior of Josephson current is entirely controlled by the ultraviolet cutoff ($\Lambda$) of the tilted Dirac theory. 
The number of channels transporting the Josephson current will also be determined by the same cutoff. Therefore a suitable quantity 
that can separate the DOS effects from the other effects is the Josephson current {\em normalized} to the total number of transmission channels.

\begin{figure}[b]
\centering
\includegraphics[width=8cm]{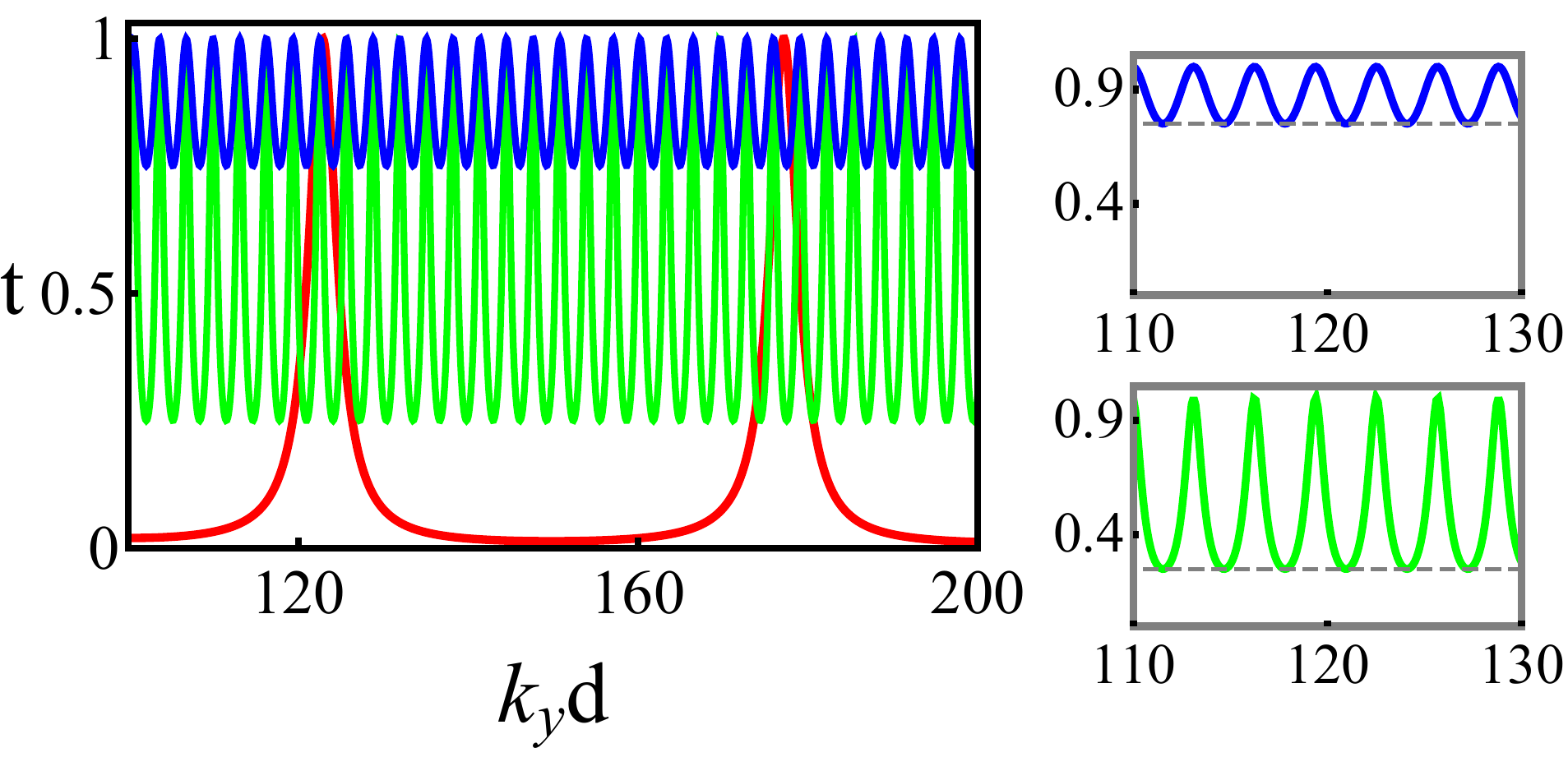}
\caption{The transmission probability for large $k_yd$ at $\zeta=1$ for three different orientation of tilt vector, $\theta=\pi/6$ (Blue), 
$\theta=\pi/3$ (Green) and $\theta=\pi/2$ (orange). Right panels enlarge the left panel for better resolution. The offset values, $t_b$ are indicated by dashed lines.  }
\label{t.fig}
\end{figure}

This quantity can be easily calculated by considering the large $k_y$ limit of the transmission probability ${\mathfrak t}$, Eq.~\eqref{t.eqn},
which gives
\bea
{\mathfrak t}=[\cos^2 (k_y d)+\sin^2 (k_y d) / \cos^2 \theta]^{-1}.
\label{tappx.eqn}
\eea
Thus  the transmission probability around the $n$'th channel  defined by $n\pi/d -x<k_x<n\pi/d + x$, 
where $x$ is a small wave vector (much smaller than $\pi/d$), can be separated into two contributions as,
\bea
{\mathfrak t}={\mathfrak t}_b+(1+x^2 \tan^2\theta )^{-1}.
\label{tpeak.eqn}
\eea
where the parameter ${\mathfrak t}_b=\cos^2 \theta$ is an offset value of  ${\mathfrak t}$.
These state of affairs are represented in Fig~\ref{t.fig}. 
Note that in calculating $I_c= e\Delta \sum {\mathfrak t}/(1- {\mathfrak t}/2)^{1/2}$, the summand is an oscillating function of $k_y$ 
that is maximized when $\mathfrak t=1$. This maximum happens in two situations:
(i) when $\mu=0$  and (ii) when $k_x^e L = n \pi$ with $n$ an integer.  
The number of channels (maximum value $N$ of the integer $n$) is determined from asymptotic limit of Eq.~\eqref{kx.eqn}
for large $k_yd$ by saturating the $k_y$ with the ultraviolet cutoff $\Lambda$,
\bea
N=\Lambda \cot \theta + \frac{\sqrt{2\Lambda |\mu| \sin\theta}}{\sin^2\theta}.
\label{N.eqn}
\eea 

\begin{figure}[t]
\centering
\includegraphics[width=7cm]{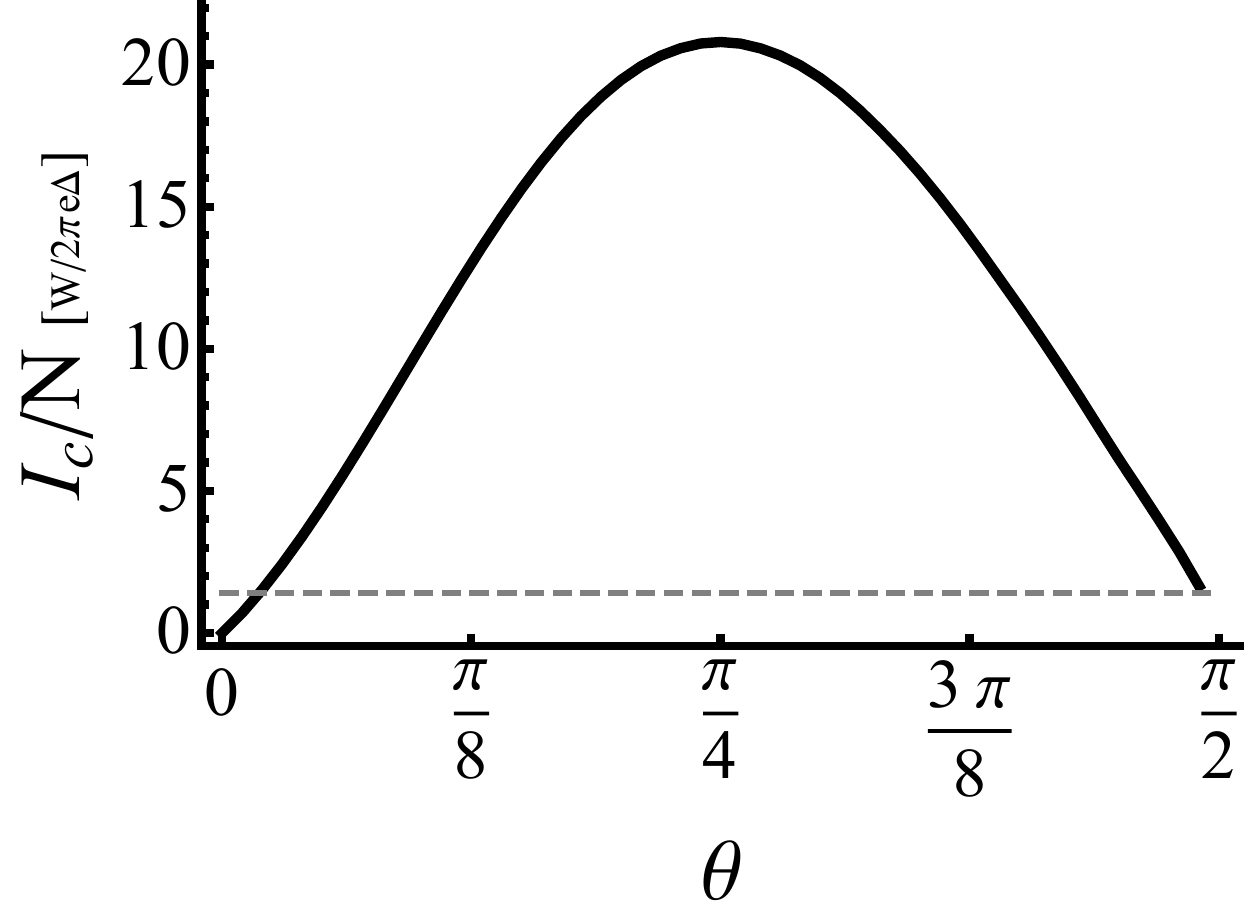}
\caption{ Critical Josephson current normalized to the total number of transmission channels as a function of 
the tilt orientation ($\theta$) for the case $\zeta=1$. $N$ is the number of propagating modes.}
\label{IN_teta.fig}
\end{figure}

Using Eq.~\eqref{tpeak.eqn} and expanding up to second order in $x$, gives the following
expansion for the $I_c$ around the peak values,
\bea
I_c^{\rm peak}(x) \approx\frac{2-3 x^2 \tan^2 \theta }{\sqrt 2},
\label{ipeak.eqn}
\eea
The integral giving rise to critical Josephson current contains two terms. One comes from the 
offset transmission ${\mathfrak t}_b$. The other contribution comes from the peaks in the oscillatory
part. Using the above approximation for the peaks, the critical Josephson current away from $\theta=0,\pi/2$ becomes
\be
   \frac{I_c}{e\Delta}\approx g(\theta) \Lambda/ 2 + N\sqrt{\frac{2}{3} g(\theta)} \frac{4+ f(\theta)}{3 \tan\theta}
\label{Icinteg.eqn}
\ee 
where $g(\theta)=\cos^2 \theta + \cos ^4 \theta$. For a non-zero $\theta$, the cutoff $\Lambda$ is proportional to
the number $N$ of the transmission channels. Therefore the above expression can be normalized to $N$. 
Fig.~\ref{IN_teta.fig} shows the normalized critical Josephson current as a function of tilt orientation.
For $\theta=0$ the oscillatory part goes away and there are no peaks. So the oscillatory contribution to
integral vanishes. In this limit, the $N$ in Eq.~\eqref{N.eqn} diverges and hence $I_c/N$ in this figure vanishes. 
But it does not mean that the $I_c$ itself is small. Because in $\theta=0$ limit, the background value ${\mathfrak t}_b=1$ is maximal. 
For $\theta=\pi/2$, the width of the peaks in approximation~\eqref{ipeak.eqn} vanishes, and also ${\mathfrak t}_b=0$. Therefore 
the plot in Fig.~\ref{IN_teta.fig} saturates to $\sqrt 2$. 
This figure separates the density of states ($N$) effect from the anisotropy caused by the 
orientation $\theta$ of the tilt vector $\vec\zeta$, and leaves a non-trivial $\theta$-dependence.
Therefore the enhancement of the critical Josephson current by approaching $\zeta\to 1$ is not a pure
density of state effect. In addition it involves a non-trivial dependence on the orientation of the tilt. 

The limit $\theta\to 0$ of Eq.~\eqref{N.eqn} deserves a further discussion:
In this limit, the second term of Eq.~\eqref{Icinteg.eqn} that estimates the area under the peaks will
be zero, as in this limit there will be no peaks and this equation reduced to
\be
   \frac{I_c}{e\Delta}\approx g(\theta) \Lambda/ 2
\label{Icinteg2.eqn}
\ee
Now let us turn our attention to Eq.~\eqref{N.eqn} and discuss the $\theta\to 0$ limit. 
When $\theta$ is away from zero, for large $\Lambda$, the first term dominates over the second term
and therefore the number $N$ of channels is controlled by the cutoff $\Lambda$. However in the
$\theta\to 0$ limit, the divergence $\theta^{-3/2}$ of the second term dominates over the $\theta^{-1}$ of the 
first term. In this limit, the relation between the cutoff $\Lambda$ and the number $N$ of the modes will be
given by 
\be
   N\propto \frac{\sqrt {\Lambda |\mu|}} {\theta^{3/2}}.
\ee
Plugging the above equation into Eq.~\eqref{Icinteg2.eqn} 
and using the fact that $g(\theta=0)=2$, for a fixed but small $\theta$ gives
\be
   \frac{I_c}{e\Delta}\approx N^2\theta^3
\ee
It is remarkable to note that at $\zeta=1$, and for $\theta\to 0$, the critical current
is proportional to $N^2$ where $N$ is the number of available channels. This behavior
is unusual as one normally expects the Josephson current to be proportiona lto $N$. 
Given that $\zeta=1$ corresponds to an event horizion in the geometric interpretation,
the $N^2$ dependence reflects a horizon peroperty. The hallmark of event-horizon is
that the "gravity" forces (in our case all forces come from Coulomb interaction of charges) are so strong
that the virtual electron-positron pairs (in our case electron-hole pairs) become on-shell
and therefore electron-hole pairs will be proliferated~\cite{CarrolBook}. This is the 
textbook explanation of the Hawking radiation~\cite{CarrolBook}. In the present case it seems 
that when the tilt is just perpendicular to the superconducting interface, the proliferated
electron-hole pairs get repeatedly Andreev reflected~\cite{Faraei_PAR}, thereby generating $N^2$ terms.

\section{Summary and outlook}
\label{discuss.sec}
In this work we studied an SNS structure based on tilted Dirac fermions. The tilt is parameterized by a vector
$\vec\zeta=(\zeta_x,\zeta_y)=\zeta(\cos\theta,\sin\theta)$. For $\zeta<1$ we studied the system in two regimes of 
long and short junctions corresponding to Thouless energy $E_T=\hbar v_F/d\ll\Delta_0$ and $E_T\gg\Delta_0$ where 
$\Delta_0$ is the s-wave superconducting gap and $d$ is the width of the normal tilted Dirac material junction. 

In the long junction limit when the chemical potential is tuned to specular Andreev reflection regime
defined by $\eps>\mu$, the propagating Andreev modes are formed along the channel ($y$ direction in Fig.~\ref{model.fig}).
Each branch of Andreev modes of the upright Dirac cone~\cite{Andreevmodesgap} have an incipient four-fold degeneracy 
which is split by the tilt vector $\vec\zeta$ as in Fig.~\ref{e_ky.fig}. 
The two colors arise from breaking $k_x\to -k_x$ symmetry of the Dirac equation by the tilt vector $\vec\zeta$, while
a "bifurcation"-like splitting is due to the valley index $\tau$. 
The important consequence of this splitting is that the semiclassical paths of Andreev modes are distorted 
in such a way (see Fig.~\ref{blured.fig}) that, unlike the upright case, a net electric charge current can be obtained 
from the Andreev modes. This manifests itself as a tilt-dependent charge current in response to thermal gradient
which would have been otherwise zero.
The distinguishing feature of the current due to Andreev modes from the current due to
normal electrons is its dependence on the phase difference $\phi$ of the two superconductors which can be detected 
by a flux bias~\cite{Inanc2019PRL}.
Other transport coefficients also receive corrections that are all odd functions of $\vec\zeta$. 

In the short junction limit where instead of propagating Andreev modes, localized Andreev levels are formed,
the tilt dependence can be nicely imprinted into the Josephson current. The first important observation is that
the Josephson current can be enhanced by orders of magnitude by bringing the tilt closer and closer to the $\zeta=1$ limit.
In a geometric language this limit corresponds to an event-horizon of the underlying metric~\cite{Volovikblackhole,tohid,Saharmetric}.
In this particular limit, the physics is particularly clear: Part of the enhancement is a density of states effect
and the resulting Josephson current is proportional to the number $N$ of the transmission channels. However, there remains
an additional dependence on the direction $\theta$ of the tilt vector shown in Fig.~\ref{IN_teta.fig}. 

In the $\theta\to 0$ limit, the Josephson current turns out to be proportional to $N^2$ (rather than proportional to $N$) 
where $N$ is the number of available channels. This counter-intuitive result appears to be a property of event-horizon (corresponding
to $\zeta=1$) which can be interpreted by a pair creation mechanism responsible for Hawking radiation.
Given the parallel between geometrical approaches and our present approach based on Landauer-formula, an explicit calculation relating 
the $N^2$ dependence to Hawking radiation is desirable.

\section{Acknowledgements}
Z. F. is grateful to the Abdus Salam center for Theoretical Physics for a long term visit during which this research was initiated.
We thank R. Fazio for discussions. 
S. A. J. was supported by grant No. G960214 from the research deputy of Sharif University of Technology
and Iran Science Elites Federation (ISEF). S. A. J. appreciates Prof. Durmus Ali Demir of Sabanci University for useful discussions
on the Hawking radiation.

\bibliography{mybib}

\end{document}